\documentclass[twocolumn,showpacs,amsmath,amssymb]{revtex4-1}
\usepackage{graphicx}% Include figure files
\usepackage{dcolumn}% Align table columns on decimal point
\usepackage{bm}% bold math
\usepackage{url}
\usepackage{amssymb}
\usepackage{wasysym}
\usepackage[stable]{footmisc}
\usepackage{slashbox}
\usepackage{longtable}
\usepackage{amsmath}
\usepackage{tabularx}
\usepackage{float}
\usepackage{mathptmx}
\usepackage{longtable}
\begin{document}

\title{Nanoscale Nonlinear Radio Frequency Properties of Bulk Nb : Origins of Extrinsic Nonlinear Effects}% Force line breaks with \\

\author{Tamin Tai$^{1,2}$}
\email{tamintai@gmail.com}
\author{B. G. Ghamsari$^{2}$}
\author{T. Bieler$^{3}$}
\author{Steven M. Anlage$^{1,2}$}
\affiliation{$^{1}$Department of Electrical and Computer
Engineering, University of Maryland, College Park, MD 20742-3285,
USA}

\affiliation{$^{2}$Department of Physics, Center for Nanophysics
and Advanced Materials, University of Maryland, College Park, MD  20742-4111, USA}

\affiliation{$^{3}$Chemical Engineering and Material, Michigan State University, East Lansing, MI 48824, USA}

\date{\today}% It is always \today, today,
\newcommand{\squeezeup}{\vspace{-2.0mm}}

\begin{abstract}
The performance of Niobium-based Superconducting Radio Frequency
(SRF) particle accelerator cavities can be sensitive to localized defects that
give rise to quenches at high accelerating gradients. In order
to identify these material defects on bulk Nb surfaces at their
operating frequency and temperature, a wide bandwidth microwave microscope with localized and strong RF magnetic fields is developed by integrating a magnetic write head into the near-field microwave microscope to enable mapping of the local electrodynamic response in the multi-GHz frequency regime at cryogenic temperatures. This magnetic writer demonstrates a localized and strong RF magnetic field on bulk Nb surface with $B_{surface}$ $ > $ $10^{2}$ $mT$ and sub-micron resolution. By measuring the nonlinear response of the superconductor, nonlinearity coming from the nano-scale weak link Josephson junctions due to the contaminated surface in the cavity fabrication process is demonstrated.
\end{abstract}

  %Use showkeys class option if keyword
\maketitle

\section{Introduction}

There is continuing interest to build high energy
accelerators to bring electrons and positrons up to near light velocity using
bulk Niobium (Nb) superconducting radio frequency (rf) cavities \cite{ILC}. These electromagnetic cavities resonate at microwave frequencies
and can efficiently transfer microwave power into the kinetic energy of the charged particle beam \cite{H. A. Schwettman}\cite{Padamsee1}. Each cavity operates with regions of high electric and magnetic rf fields, with the highest electric field on the accelerating axis and maximum magnetic field on the equator surface of the cavity \cite{Padamsee1}.
However to put these lab-fabricated cavities into commercial
mass-production, many issues arise. It is a challenge to manufacture such high performance
cavities with a high degree of consistency. Much debate regarding this inconsistent performance
concluded that certain types of defects on the Nb cavity surface can
behave as a source of quenching in tangential high RF magnetic fields \cite{Ciovati2010}\cite{Sung2011}\cite{Polyanskii}. A quench is when the Nb superconductor returns to the normal state in the presence of strong fields, thus limiting their utility. Candidates for these defects include pits, oxides, hydrides, impurity inclusions, grain boundaries etc. \cite{SRF_tutorial2011}\cite{Sung2014}. These defects arise from cavity fabrication steps such as forming, machining, electron beam welding, electro polishing (EP) and buffered chemical polishing (BCP). These defects are either non-superconducting
or have lower $T_c$ (for example, NbO$_x$ with $x=0.02 \sim 0.04$ has $T_c$ ranging from 5.1 K to 6 K \cite{HalbritterSRF}), making them sources of dissipation and interrupting the superconducting current flow. Unfortunately, it is difficult to totally remove all of these defects even after sophisticated physical and chemical treatments. Because different defects have their
individual quench limit, not all of the defects behave as sources of
quenching in certain operation conditions (for example, a 9 cell TESLA cavity
usually operates at a temperature of 2 K, an initial accelerating gradient of 25 MV/m, 1.3
GHz RF frequency under 16.5 MW microwave power \cite{Padamsee1}).
Unfortunately, it is not known which defects have lower quench limits and therefore it is necessary to quantitatively understand each type of defect.

\begin{figure*}
\begin{quote}
\centering
\includegraphics[width=4.5in]{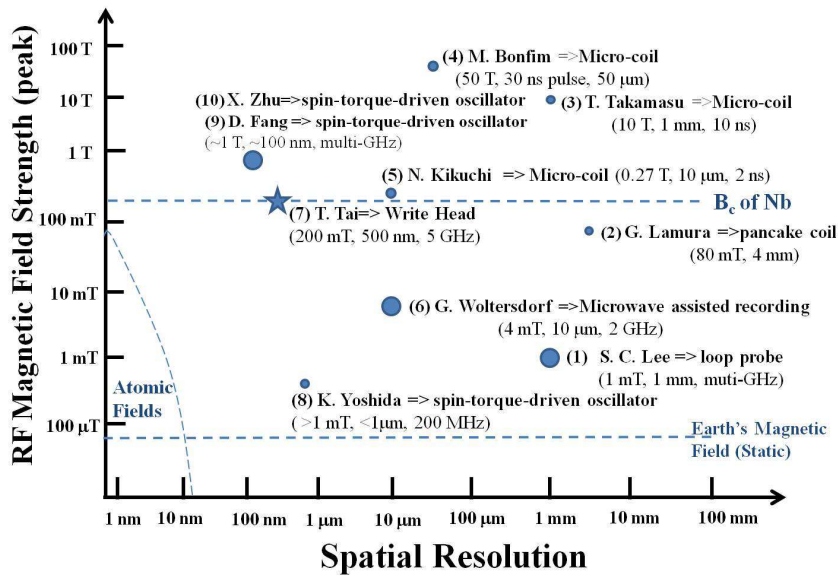}
\caption {High-Field, High-Resolution AC/Transient Magnetic Microscopy Generating Highly-Confined Magnetic Fields with Frequency $\geq$ 100 MHz : A selected summary of methods to generate a highly concentrated RF magnetic field. Each technique lists a typical RF magnetic field scale, degree of spatial confinement/resolution, and an associated time/frequency scale demonstrated. All of the candidate technologies are numbered and referenced as follows
(1) S. C. Lee $\Rightarrow$ loop probe. \cite{S. C. Lee1} \cite{S. C. Lee2} \cite{S. C. Lee3}.  (2) G. Lamura  $\Rightarrow$ pancake coil \cite{Lamura} (3) T. Takamasu $\Rightarrow$ micro-coil \cite{Takamasu} (4) M. Bonfim $\Rightarrow$ micro-coil \cite{Bonfin} (5)N. Kikuchi  $\Rightarrow$ micro-coil \cite{Kikuchi} (6) G. Woltersdorf  $\Rightarrow$ microwave assisted  recording \cite{Woltersdorf}. (7) T. Tai $\Rightarrow$ magnetic write head \cite{Tai2013}\cite{Tai2014_APL}. (8) K. Yoshida $\Rightarrow$ spin-torque-driven oscillator \cite{Yoshida}. (9) D. Fang $\Rightarrow$ spin-torque-driven oscillator \cite{Fang}. (10) X. Zhu $\Rightarrow$ spin-torque-driven oscillator \cite{Zhu}. Note that larger circle $(\newmoon)$ and star sign $(\bigstar)$ indicates the operation frequency/speed corresponds to higher frequencies. The star sign also indicates the method used in this paper.}
\label{Chapter1_summary}
\end{quote}
\end{figure*}

Due to the defect issues with Nb, ideally one would like a microscopic
technique that identifies defects based on their poor microwave
performance at low temperatures in the superconductive state.
Quenches often occur in regions of the cavity with strong tangential magnetic field, hence a local probe with strong rf magnetic field concentration is desired.
Generally speaking, creating a concentrated microwave magnetic field in a small area of the sample is a challenge and has created interest among several research groups using different methods. Fig. \ref{Chapter1_summary} shows a brief summary of developments of localized high RF magnetic field measurements. Our objective is to move into the upper-left quadrant of this diagram, in other words to develop a microscopic probe that simultaneously produces a very strong rf magnetic field over a very limited area of the sample. The conventional method is to make a loop-shaped coil as small as possible to generate a strong magnetic field on the sample \cite{S. C. Lee1}\cite{S. C. Lee2}\cite{S. C. Lee3}\cite{Mircea1}\cite{Clinton}. Some groups also bundle many loop-shaped coils to generate a strong localized field \cite{Lamura}, but unfortunately this method will sacrifice spatial resolution. Recently several groups have used the method of either photo-lithography or E-beam lithography to fabricate a micro-coil and successfully generated about 10 to 50 Tesla of pulsed magnetic field on 30 ns time scales \cite{Takamasu}\cite{Bonfin}. One group also demonstrated that this nanosecond pulsed field can switch the magnetization of nano magnetic particles \cite{Kikuchi}, but this method will generate heat and may not be useful to probe the properties of superconductors due to the complicated sample heating during the measurement. Another study on utilizing spin-torque-driven microwaves to generate a localized RF magnetic field to flip the magnetization direction of a thin single domain magnetic element has been widely discussed recently \cite{Woltersdorf}\cite{Yoshida}\cite{Fang}\cite{Zhu}\cite{Sim}. However this technique doesn't have application yet in investigating the properties of SRF cavities.

In the microwave microscope approach described here, a magnetic write head from a hard disk drive is used as a near field scanning probe to generate strong and localized RF field on a superconducting sample \cite{Tai2011}\cite{Tai2012}\cite{Tai2013}\cite{Tai2014_JAP}\cite{Tai2014_APL}. The device is a high resolution magnetic microscope probe in the GHz frequency regime. The superconducting
sample responds by creating screening currents to maintain the Meissner state in
the material. These currents inevitably produce a time-dependent variation in the
local value of the superfluid density, and this in turn will generate a response at
harmonics of the driving tone. At localized defective regions, which will cause pre-mature quench of the cavity, the enhanced harmonic response will be more significant.

\begin{figure}
   \centering
   \includegraphics*[width=3.5 in]{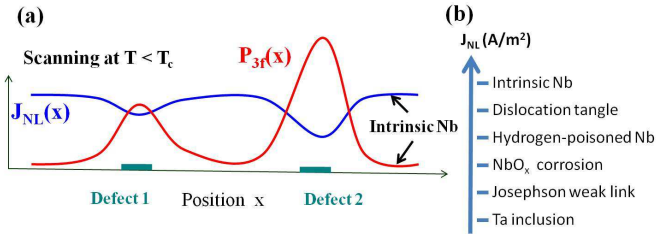}
   \begin{quote}
   \caption{(a) Schematic $J_{NL}$ variation on the superconducting surface from intrinsic Nb to different discrete defects. The $P_{3f}$ generation from defects is local. The defects with greatest disruption of the superfluid density have the lowest $J_{NL}$ and strongest $P_{3f}$. Hence measuring $P_{3f}$ at different positions can be used to find defects on superconductors. (b) A schematic hierarchy of a few representative defects and their associated $J_{NL}$ values. }
   \label{JNLP3f}
   \end{quote}
   \squeezeup
\end{figure}

The nonlinear third-harmonic response frequency is studied in this paper because that one is intrinsically always present (to greater or lesser extent) in a superconductor in zero dc field at finite temperature.  In addition to this intrinsic response, extrinsic properties (i.e. associated with defects) generate their own third harmonic response, almost always at a level far stronger than the intrinsic response \cite{Oates2007}. Figure \ref{JNLP3f} (a) illustrates a schematic situation in which defects stimulated by a localized scanned RF magnetic field probe show enhanced nonlinear response compared to the background intrinsic response.
One can parameterize the nonlinear response with an associated ``nonlinear critical current density" called $J_{NL}$. The nonlinearity is quantified in terms of the effect that the defect has on the superfluid density $n_s$ with increasing RF current $J$. This is written as $\frac{n_s (T,J)}{n_s (T,0)} \cong 1-(\frac{J}{J_{NL}})^2$ , showing that regions with small $J_{NL}$ will have enhanced suppression of the superfluid density \cite{S. C. Lee1}\cite{S. C. Lee2}\cite{S. C. Lee3}\cite{Booth1999}. The values of $J_{NL}$ associated with the intrinsic Nb background material, and several defects, are also shown schematically in Fig. \ref{JNLP3f} (b). The third harmonic power is inversely proportional to $J_{NL}$ as $P_{3f} \sim (\frac{1}{J_{NL}})^4$ in thin film superconductors \cite{S. C. Lee1}\cite{S. C. Lee2}\cite{S. C. Lee3}. Hence defective regions with smaller $J_{NL}$ will produce larger harmonic power $(P_{3f})$, revealing their presence in the scanned probe microscope.

This paper will first address experimental measurements (section $\mathbf{II.}$ and section $\mathbf{III.}$) of the nonlinear third harmonic response from bulk Nb materials in localized regions down to sub-micron length scales. In section $\mathbf{IV}$, a detailed discussion of the quantitative modeling of the experimental results will be given and then section $\mathbf{V}$ contains the conclusion. This paper provides a more detailed examination of the preliminary data presented in Ref. \cite{Tai2014_APL}.

\section{Experiment}
\begin{figure}
\centering
\includegraphics*[width=2.3in]{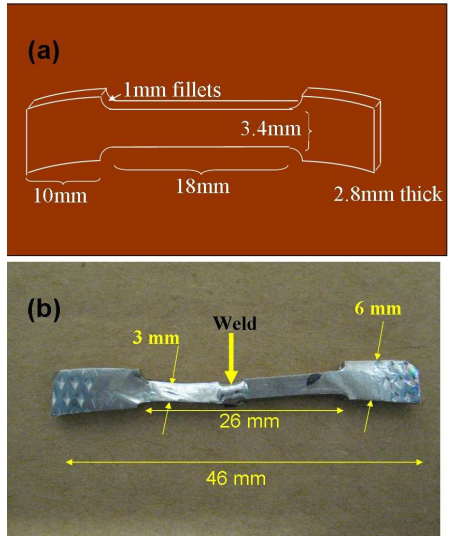}
\begin{quote}
\caption{(a) A schematic picture of the bulk Nb tensile specimen before
deformation. All dimensions of the specimen are labeled. (b) The
finished specimen after tensile test and welding treatment in the
center of the sample.} \label{Bulk_Nb}
\end{quote}
\squeezeup
\end {figure}

\begin{figure}
\centering
\includegraphics*[width=3.5in]{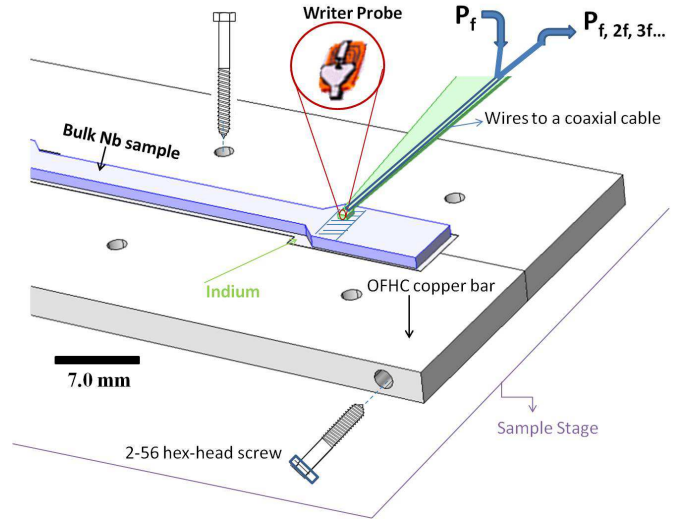}
\begin{quote}
\caption{The setup inside the cryostat probe station for measuring the bulk Nb nonlinearity. The slashed region on the top of the bulk Nb sample indicates where the $P_{3f}$ measurements are performed. The circled zoom-in picture schematically highlights the magnetic write head probe on the surface of the bulk Nb. Note that thermal anchoring is critical for cooling the bulk Nb sample below $T_c$.}
\label{Bulk_Nb_Setup}
\end{quote}
\squeezeup
\end {figure}

A bulk Nb sample with $T_c$ of 9.2 K and well-defined locations of defects is an ideal candidate for measurement of nonlinearity using a microwave microscope.
Fig. \ref{Bulk_Nb} (a) shows a schematic diagram of the original tensile specimen made from bulk Nb with a high residual-resistivity ratio (RRR). The surface of three tensile specimens were mechanically polished and then electropolished prior to deformation. To simulate an elliptical SRF cavity forming a welding process using large grain niobium, single crystal samples were strained to $40\%$ elongation, cut in half, and different halves were welded together.
Fig. \ref{Bulk_Nb} (b) shows the finished specimen. The sample has been carefully
imaged by electron backscattered pattern (EBSP) orientation imaging microscopy (OIM) \cite{Bieler} and a variety of microstructural conditions were identified, including as-deformed regions that were kept cold near the end, recrystallized regions with several grain boundaries in the center, and recovered regions in between, many niobium oxides, grain boundaries and dislocations on the surface. The OIM analysis identified several specific defects that appear in the
processing of bulk Nb cavities.

In order to make the sample surface superconducting in our microscope
cryostat system, thermally anchoring the sample and good radiation
shielding from the outside environment is crucial. Figure \ref{Bulk_Nb_Setup} shows the setup inside the cryostat. The bulk Nb sample is clamped in place by two halves of an oxygen-free high conductivity (OFHC) copper bar. Indium is filled within the gap between the Nb sample and OFHC copper bar to reduce the thermal resistance between bulk Nb and the OFHC copper bar. The two halves of the bar are bolted together laterally by two $2-56$ hex-head screws from the side and then the whole OFHC copper bar assembly is bolted directly onto the surface of the sample stage in the cryostat with vacuum grease in between to reduce the thermal resistance. Note that the sample stage is also made of OFHC copper and is also firmly bolted on the top of the cold plate. A thermometer is placed on the bulk Nb top surface to monitor the temperature of the relevant surface of the sample. An OFHC copper cryogenic radiation shield is attached to the cold head of the cryostat. The lid of the radiation shield has a 3 inch diameter infrared blocking window as a view port. This setup can cool down the bulk Nb sample top surface to 5.0 K without pumping the exhaust from the cryostat. By pumping the exhaust, the surface temperature of the bulk Nb can go below 4.0 K. However 5.0 K is cold enough to analyze the surface microwave properties of this bulk Nb sample. Note that if these extensive precautions are not taken, it is virtually impossible to cool the surface of bulk Nb down below its transition temperature. More photos inside the vacuum chamber can be found in the reference \cite{Tai_thesis}.

The principles of measuring the complex third harmonic voltage $(V^{sample}_{3f})$
or related scalar power $(P^{sample}_{3f})$ that is generated by the sample due to local excitation at a fundamental frequency from the magnetic write head has been widely discussed in previous publications \cite{Tai2011}\cite{Tai2012}\cite{Tai2013}\cite{Tai2014_APL}. The magnetic write head is made by Seagate (part $\sharp$ GT5) and was originally designed for longitudinal magnetic recording. Hence it creates a strong tangential RF magnetic field on the surface of the sample. The superconducting sample in proximity to the probe responds by creating screening currents to maintain the Meissner state in the material. These currents inevitably produce a time-dependent variation in the local value of the superfluid density, and this in turn will generate a response at harmonics of the driving tone. A high pass filter is used outside the cryostat to suppress the reflected $P_{f,2f}$ and pick up the $P_{3f}$ signal.

\section{Third Harmonic Measurement Results on Bulk Nb}

\begin{figure}
\centering
\includegraphics*[width=3.0in]{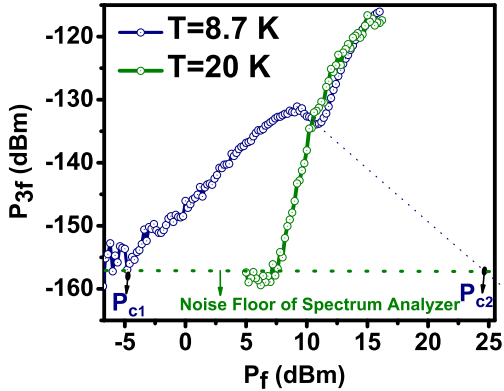}
\begin{quote}
\caption{The $P_{3f}$ dependence on $P_f$ at a selected temperature, 8.7 K, for local microwave excitation of bulk Nb with the magnetic write head probe. Note the $T_c$ of the bulk Nb sample is 9.2 K. The $P_{3f}$ at $T$=20K results from the probe nonlinearity itself.}
\label{NbP3fPfSet1}
\squeezeup
\end{quote}
\end{figure}
Many localized third harmonic measurements are performed in the slashed region indicated in Fig. \ref{Bulk_Nb_Setup}. Based on OIM images, this region has a moderately high density of disorganized clusters or knots of dislocations due to the presence of some plastic deformation from the tensile test. The probe height is determined by optical microscopy from the side of the cryostat combined with $S_{11}(f)$ measurement of the magnetic write head to decide the best excitation frequency. The probe height is approximately within 1 $\mu m$ from the bulk Nb surface. The details of the probe height control are discussed in Ref. \cite{Tai2014_JAP}\cite{Tai_thesis}. Different locations show slightly different microwave nonlinear properties and will be illustrated qualitatively in this section.

\begin{figure}
\centering
\includegraphics*[width=3.0in]{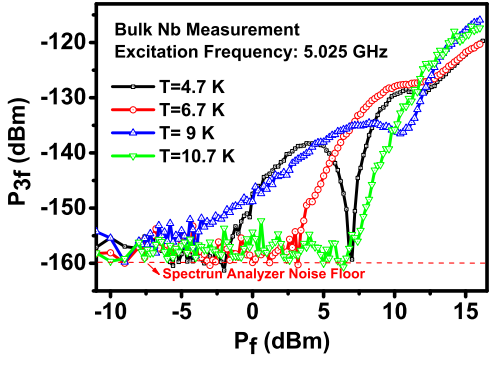}
\begin{quote}
\caption{ The $P_{3f}$ dependence on $P_f$ at selected temperatures on the bulk Nb sample with the magnetic write head probe. The $P_{3f}$ dependence on $P_f$ at 4.7 K shows a different nonlinear behavior compared to the other two temperatures. The nonlinearity of the probe is measured at T=10.7 K in this data set.}
\label{NbP3fPfSet2}
\end{quote}
\squeezeup
\end{figure}

Figure \ref{NbP3fPfSet1} shows one of the representative measurements for the power dependence of $P_{3f}$ with respect to fundamental input power $P_f$ at two fixed temperatures (above $T_c$ and below $T_c$) under 5.025 GHz excitation at this area. The representative curve measured at $T=8.7$ $K$ (below $T_c$) shows a distinct $P_{3f}$ onset from the noise floor of the spectrum analyzer followed by a continuous increase of nonlinearity until a turnover occurs at high excitation power. Then the $P_{3f}(P_f)$ data at high excitation power eventually approaches and oscillates around the probe background nonlinearity, $P_{3f}^{probe}(P_f)$, measured at $T=20 K$. Firstly, the onset of the nonlinear response can be defined as a temperature dependent lower critical power, $P_{c1}$. This onset power, $P_{c1}$, will shift toward larger excitation power at lower temperature on the Nb superconductor (Fig. \ref{NbP3fPfSet2}) \cite{Tai2014_APL}. In addition, after a turnover, all $P_{3f}(P_f)$ curves tend to decrease with increasing power, suggesting that Nb superconductivity is suppressed. The linear extrapolation of $P_{3f}$ to the noise floor of the spectrum analyzer can be defined as an upper critical power, $P_{c2}$, which suggests that superconductivity will eventually be destroyed due to the high RF magnetic field. It is seen that higher $P_{c2}$ is required to destroy the superconductivity at lower temperature \cite{Tai2014_APL}. The temperature dependent $P_{c1}$ and $P_{c2}$ can be associated with temperature dependent surface field $B_{c1}$ and $B_{c2}$ (or perhaps $B_{c}$).  The analysis of the temperature dependent $P_{c2}$ has been used to estimate the surface field excited by the magnetic write head probe. This analysis has been carried out by a fit of the experimental temperature dependent critical field $B_{c}(T)$ from each experimental $P_{c2}$ with the approximate equation $B_{c}(T) \cong B_{c}(0K) \Big(1-(T/T_c)^2\Big)$, valid near $T_c$ \cite{Tai2014_APL}. The fit indicates localized magnetic field from the magnetic write head probe is on the order of the thermodynamic critical field of Nb with B$_{surface}$ $\sim 10^2 mT$ \cite{Tai2014_APL}. Different positions will show different $P_{c1}$ and $P_{c2}$ values due to the inhomogeneity of the bulk Nb surface properties.

If the inhomogeneity of the Nb surface contains Nb hydride or Nb oxide, which has lower critical temperature there may form a weak link Josephson junction. In this case, the curve of $P_{3f}$ versus $P_f$ will behave different from the case of the homogeneous bulk. Fig. \ref{NbP3fPfSet2} shows the $P_{3f}$ versus $P_f$ at 4.7 K comparing to that at T=9 K and T=6.7 K. One can see at T=4.7 K, the curve of $P_{3f}$ versus $P_f$ shows an additional peak while the excitation power is between -2.5 dBm and 7 dBm. This implies some additional source of nonlinearity is excited at this temperature. Detailed interpretation will be given in the section IV.

\begin{figure}
\centering
\includegraphics*[width=3.0in]{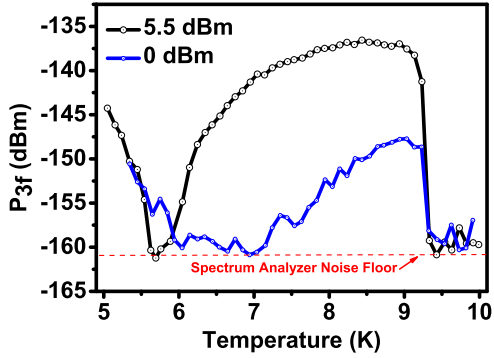}
\begin{quote}
\caption{The temperature dependent $P_{3f}(T)$ (measured as temperature decreases) of the bulk Nb sample at two specific input powers and 5.025 GHz excitation. Note the $T_c$ of the bulk Nb sample is 9.2 K and the onset of nonlinearity at low excitation power also begins at exactly 9.2 K.}
\label{BulkNbP3fTset2}
\end{quote}
\squeezeup
\end{figure}

The temperature dependence of $P_{3f}(T)$ at the same probe position is shown in Fig. \ref{BulkNbP3fTset2} at two values of $P_f$. The onset of Nb nonlinearity begins immediately at the bulk Nb $T_c$ (9.2 K) and then quickly saturates before T=8 K for $P_f$=5.5 dBm, 5.025 GHz excitation. The nonlinearity shows a dip around 5.8 K. Below 5.8 K, the nonlinearity increases again. This same qualitative behavior also happens at a lower excitation power of $P_f$=0 dBm. From this temperature dependent nonlinear behavior, the measured response at high power is consistent with the nonlinearity from the Josephson effect \cite{Carson Jeffries}\cite{Xia and Stround}. This situation will happen on the bulk Nb surface due to the presence of oxides and hydrides forming weak link Josephson junctions \cite{Halbritter}\cite{Ciovati2006}, for example.
In the next section, the nonlinear mechanisms active in the bulk Nb measurement are modeled by the weak-link Josephson nonlinear effect.

\section{Modeling nonlinearity}
Possible extrinsic nonlinear mechanisms for the bulk Nb measurements includes moving vortices \cite{Gurevich2008} and the weak-link Josephson effect \cite{Carson Jeffries}\cite{Xia and Stround}.
The data presented in section $\mathbf{III}$ can be reasonably well explained by the nonlinear electrodynamics of the bulk Nb superconductor suffering phase slips \cite{Tinkham_Book} across a Josephson junction \cite{Xia and Stround}. Josephson junctions will form on the inner surface of the bulk Nb cavities due to the inevitable oxidation in air \cite{Ciovati2006} and exposure to water vapor. This surface oxidation deteriorates the superconductivity because of the formation of superconductor-insulator-superconductor (S-I-S), superconductor-normal metal-superconductor (S-N-S) or superconductor-constriction-superconductor (S-C-S) structures \cite{Halbritter}. In addition, Josephson junctions may also develop on the Nb surface due to the chemical surface treatment during fabrication.

\begin{figure}
    \centering
    \includegraphics*[width=2.5in]{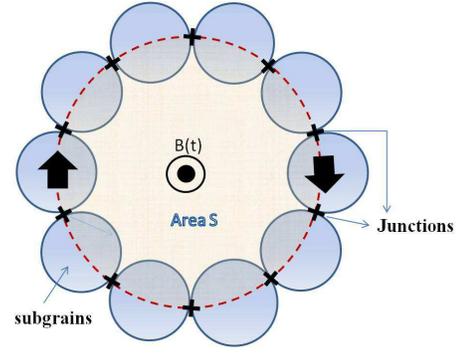}
    \begin{quote}
    \caption{Schematic illustration of an induced screening current loop created by the magnetic field pointing out of the paper. Multiple defects such as subgrains are connected by weak link Josephson junctions between the defects. Note $S$ is the area of the screening current loop. The arrow directions illustrate the direction of diamagnetic current. This figure indicates 10 Josephson junctions.}
    \label{Nb_grain_cartoons}
    \end{quote}
    \squeezeup
\end{figure}

\begin{figure}
    \centering
    \includegraphics*[width=3.2in]{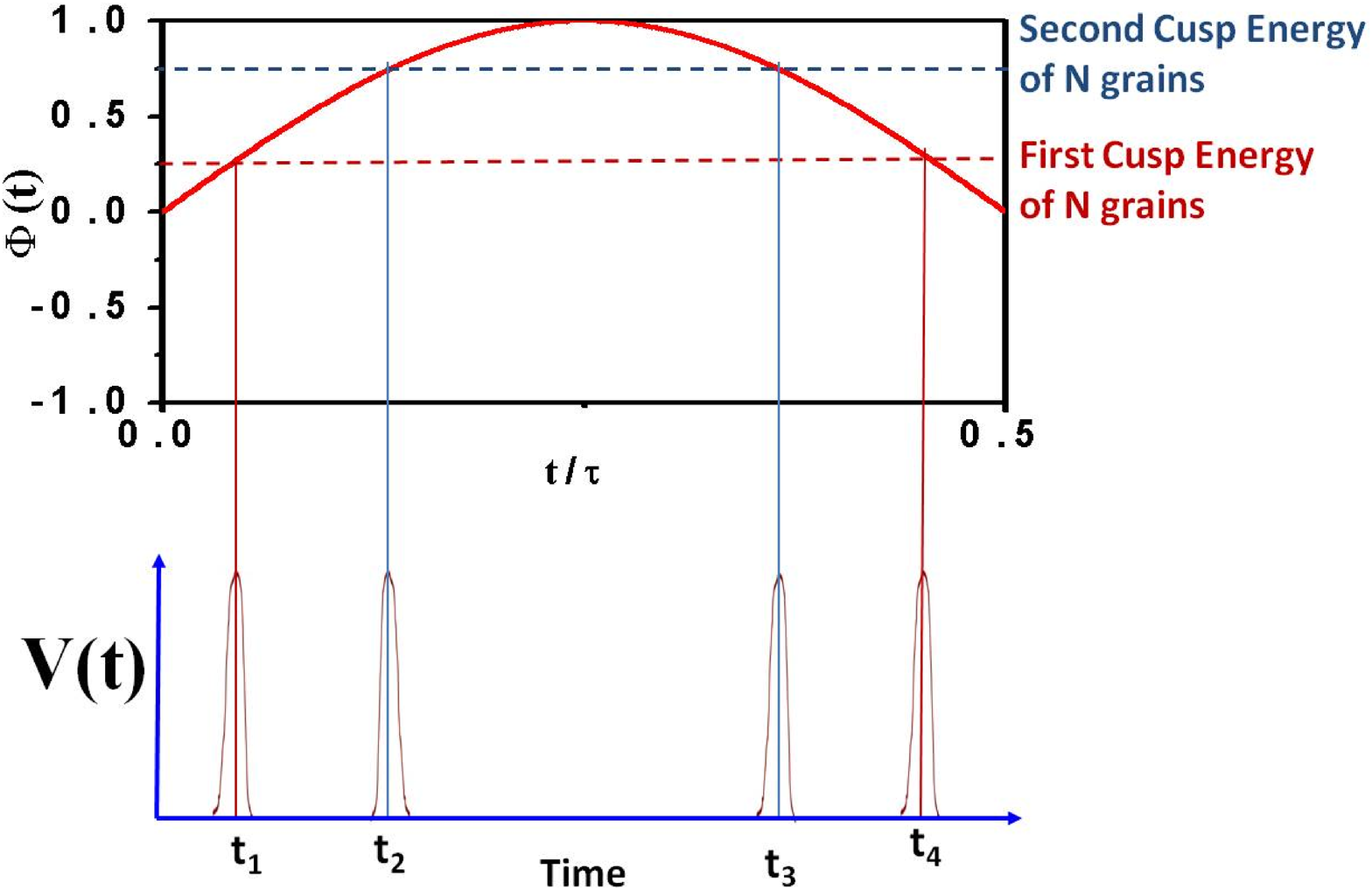}
    \begin{quote}
    \caption{(Upper frame) Normalized magnetic flux without the presence of DC field ($\Phi(t)=\Phi_{rf} sin (\omega t)$).  This figure assumes the cusps of the energy appear when $\Phi=0.25$ and $0.75$. (Bottom frame) The generated voltage spikes in a half of RF cycle ($\tau$) due to the presence of phase slips of Josephson junctions in the loop of N subgrains.}
    \label{Internal_E}
    \end{quote}
    \squeezeup
\end{figure}

To model the weak-link Josephson nonlinearity, we consider a perpendicular magnetic field component that induces a single screening current loop passing through N identical defects, which are arrangements of dislocations, dislocation tangles or impurities as shown schematically in Fig. \ref{Nb_grain_cartoons}. These sub-grains are weakly connected together in a closed loop of area $S$. This loop can be driven by a combined DC field ($B_{dc}$) and RF field ($B_{rf}$) on the superconducting surface giving rise to a total flux of
\begin{equation}\label{Phirf}
\begin{split}
\Phi(t)=\Phi_{dc}+\Phi_{rf} sin (\omega t) \\
with \qquad \Phi_{dc}=\frac{SB_{dc}}{\Phi_0}  \qquad and \qquad \Phi_{rf}=\frac{SB_{rf}}{\Phi_0}
\end{split}
\end{equation}
where $\Phi(t)$ is the total normalized magnetic flux through the loop, in units of the flux quantum $\Phi_0$, $\Phi_{dc}$ is the normalized DC flux. $\Phi_{rf}$ and $B_{rf}$ are the normalized flux and amplitude of the normal surface RF field, respectively. In this experiment, $\Phi_{dc}$ is assumed to be zero because there is no applied DC field and we assume that there is no stray DC field from the write head or the environment.
A loop of N defects under the applied RF magnetic flux ($\Phi_{rf}$) will have an internal energy ($E$) which can be defined by the Josephson coupling Hamiltonian \cite{Xia and Stround}. Clearly, if $\Phi_{rf}$ is higher than the cusp of Josephson coupling energy (either higher than the first energy cusp, the second or even higher energy cusp), a phase slip will occur in the junction \cite{Xia and Stround}. Hence, for the amplitude of RF flux shown in Fig. \ref{Internal_E}, a phase slip will occur at 4 times in the first half of the RF cycle, at times $t_1$, $t_2$, $t_3$ and $t_4,$ where $t_1$, $t_4$, are the times when the applied magnetic flux is equivalent to the first cusp of the energy of the N grain junction ring and $t_2$ and $t_3$ are the times when the applied magnetic flux is equivalent to the second cusp of the energy of the N grains.

The response voltage due to the phase slip of the Josephson junction is given by
\begin{equation}
V(t)=-\frac{\hbar}{2e} \frac{d \phi}{dt}
\end{equation}
where $\phi$ is the gauge-invariant phase difference across a Josephson junction. This voltage is a series of 4 spikes
in each half-RF period, corresponding to each of the 4 phase slip events.  We project out the
third harmonic voltage by making a sine and cosine expansion in harmonics of the fundamental
drive frequency.
\begin{equation}
\begin{split}
V_{3fa}=\frac{2}{\tau} \int_{0}^{\tau} V(t) sin(3 \omega t) dt  \\
V_{3fb}=\frac{2}{\tau} \int_{0}^{\tau} V(t) cos(3 \omega t) dt
\end{split}
\end{equation}
where $\tau$ is the rf period. This yields the following expression for the $3^{rd}$ harmonic voltage for the case of n times phase-slip events in each of period:
\begin{equation}\label{V_3f}
\begin{split}
V_{3fa}=-\frac{\hbar}{e} \frac{2 \pi \omega}{N} \Big[\sum_{i}^{n} \mid sin(3 \omega t_{i})\mid\Big] \\
V_{3fb}=-\frac{\hbar}{e} \frac{2 \pi \omega}{N} \Big[\sum_{i}^{n} \mid cos(3 \omega t_{i})\mid\Big]
\end{split}
\end{equation}
where N is the number of equivalent Josephson junctions in the loop.

\begin{figure}
    \centering
    \includegraphics*[width=3.0in]{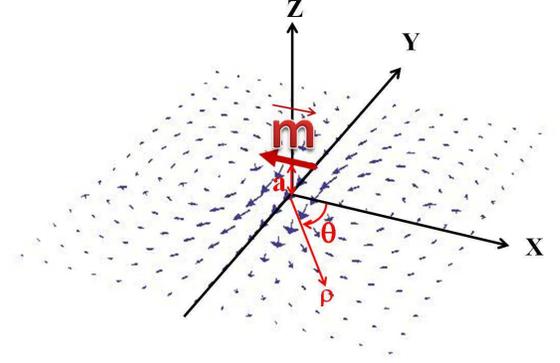}
    \begin{quote}
    \caption{The screening current (J) induced in a superconductor due to a point
    magnetic dipole $\protect\overrightarrow{m}$ with a distance $``a"$ above the superconducting surface plane $(Z=a)$. $\rho$ and $\theta$ are the radial distance and azimuth angle respectively. Surface current calculation from Ref. \cite{Peeters}. }
    \label{dioplecalculation}
    \end{quote}
    \squeezeup
\end{figure}

In the experiment, the current loop couples flux back to the magnetic write head probe. Hence what we measure is an induced voltage in the magnetic write head probe given by $V_{e}=d {\Phi_{e}}/ dt$ where $\Phi_{e}$ is the time dependent induced flux at the write head probe. To estimate the voltage induced on the magnetic probe, one can consider a magnetic dipole moment over the superconductor at a distance $``a"$ from the superconducting surface \cite{Peeters2004}\cite{Peeters2011}\cite{Bael}\cite{Peeters} as shown in Fig. \ref{dioplecalculation}. The magnetic dipole can be described by a Dirac delta function $\mathbf{M}\delta (x) \delta(y) \delta(z-a)$, where $\mathbf{M}$ is the moment of the dipole. The superconductor maintains itself in the Meissner state and generates a screening current on the superconducting surface. The radial and azimuth angle components of the generated screening current due to this dipole can be expressed as \cite{Peeters}
\begin{equation}
\begin{split}
J_{\rho}(\rho, \theta)=-\frac{ M sin \theta }{4 \pi \lambda^2} \frac{\sqrt{\rho^2+a^2}-a}{\rho^2 \sqrt{\rho^2+a^2}} \\
J_{\theta}(\rho, \theta)=\frac{ M cos \theta }{4 \pi \lambda^2} [ \frac{\sqrt{\rho^2+a^2}-a}{\rho^2 \sqrt{\rho^2+a^2}}-\frac{a}{(\rho^2+a^2)^{3/2}}]
\end{split}
\end{equation}
where $\rho$ and $\theta$ indicate the radial distance and the azimuth angle with respect to an origin on the surface just below the dipole (see Fig. \ref{dioplecalculation}). Hence, the maximum current, $J_{max}$, induced by a magnetic dipole moment $M$ at a height $``a"$ above the superconductor with penetration depth $\lambda$ will be at the location of $\rho=0$ and $\theta=-\pi/2$.
\begin{equation} \label{J_dipole}
\begin{split}
J_{max} \equiv J_{\rho} (\rho = 0, \theta = -\pi/2)=\frac{M}{8 \pi \lambda^2 a^2} \\
J_{\theta}(\rho = 0, \theta = -\pi/2)=0
\end{split}
\end{equation}

The magnetic dipole moment $M$ also can be expressed in terms of the magnetic flux in the gap of the probe as $M=\Phi_{e} l_{gap} / \mu_0$, where $l_{gap}$ is the length of the magnetic gap on the bottom of the magnetic yoke \cite{Tai2014_JAP}. Hence the induced voltage on the probe can be written as
\begin{equation}
V_{e}=\frac{d \Phi_e}{dt}=\frac{d}{dt} \frac{\mu_0 M}{l_{gap}}=\frac{dJ}{dt}\frac{8 \pi \mu_0 \lambda^2 a^2}{l_{gap}}
\end{equation}
Take an estimate of the $3^{rd}$ harmonic current density to be $J_{3}$=$I_{3}$/($\lambda$ a), where $I_{3}$ is the third order harmonic current, and in combination with Eq.\ref{V_3f}, the induced third harmonic voltage for sine and cosine expansion and the corresponding third harmonic power generated by this phase slip mechanism in the weak-link Josephson junction is given in the frequency domain by
\begin{equation} \label{V3fe}
\begin{split}
V_{3fa}^{e}= \frac{96 \mu_0 \omega \lambda(T) a I_{3}(T) }{N l_{gap}} \Big[\sum_{i}^{n} \mid sin(3 \omega t_{i})\mid \Big]\\
V_{3fb}^{e}= \frac{96 \mu_0 \omega \lambda(T) a I_{3}(T) }{N l_{gap}} \Big[\sum_{i}^{n} \mid cos(3 \omega t_{i})\mid\Big]\\
P_{3f}=\frac{\sqrt{\Big(V_{3fa}^{e}\Big)^2+\Big(V_{3fb}^{e}\Big)^2}}{2 Z_0}  \qquad \qquad \qquad
\end{split}
\end{equation}
where $Z_0$=50 $\Omega$, the characteristic impedance of the microwave generator. A complete calculation of the temperature and power dependent $I_3$ has been carried out in reference \cite{Carson Jeffries}. The outcome shows that $I_{3}=I_c(T) J_3(\beta) sin(3 \omega t)$ with $\beta\equiv \frac{2 \pi S B_{rf}}{\Phi_0} $ coming from the third order expansion of the Fourier series for $I(t)=I_c sin(\beta sin(3 \omega t))$ in terms of the Bessel function $J_n(\beta)$ \cite{Carson Jeffries}. Here $I_c$ is the junction critical current, which we take to be the Ambegaokar and Baratoff estimation \cite{Ambegaokar and Baratoff} \cite{Tinkham_Book},
\begin{equation}
I_c(T)=\frac{\pi \Delta(T)}{2eR_n} tanh[\frac{\Delta(T)}{2k_{B}T}]
\end{equation}
where $\Delta(T)$ is the temperature dependent superconducting gap parameter, and $R_n$ is the normal resistance of the junction.
The temperature dependence of $\lambda(T)$ can be calculated according to BCS theory with the zero temperature London penetration depth of Nb $\lambda_0 = 40 \; nm$ \cite{Waldram}.

To simplify the calculation for fitting experimental $P_{3f}(P_f)$ and $P_{3f}(T)$ data, the number of Josephson junctions $N$ as shown in Fig. \ref{Nb_grain_cartoons} in the current loop is assumed to be ten in Eq. \ref{V3fe}. This assumption is based on Halbritter's estimation that weak links on the surface of bulk Nb are about $10 \sim 100$ $nm$ apart \cite{Halbritter1999}\cite{HalbritterSRF}. In addition, both square bracketed terms in Eq. \ref{V3fe} are assumed to be on the order of one. The distance between the probe and sample, $a$, is assumed to be 100 $nm$, which is the scale of the magnetic gap and flying height during the reading and writing process in high speed magnetic recording \cite{Klaas B. Klaassen} \cite{ReadRiteCorp}. Two parameters, $R_n$ and $\beta$, are left for fitting.

From the power dependence measurement shown in Fig. \ref{NbP3fPfSet1} and Fig. \ref{NbP3fPfSet2}, one can see that the probe nonlinearity, $P_{3f}^{probe}$ participates in the measurement at high excitation power. In order to extract the response from only the superconducting sample at high power, an effective $P_{3f}$ is defined by taking the absolute value of the difference between the measured $P_{3f}$ in the superconducting state and the measured $P_{3f}$ in the normal state, without considering the relative phases of the contributing nonlinearity. This crude subtraction process provides a qualitative measure of the high-power nonlinear response of the sample. Therefore, the effective $P_{3f}$ can be regarded as the nonlinearity arising from just the superconducting sample, to a first approximation.
\begin{figure}
    \centering
    \includegraphics*[width=3.0in]{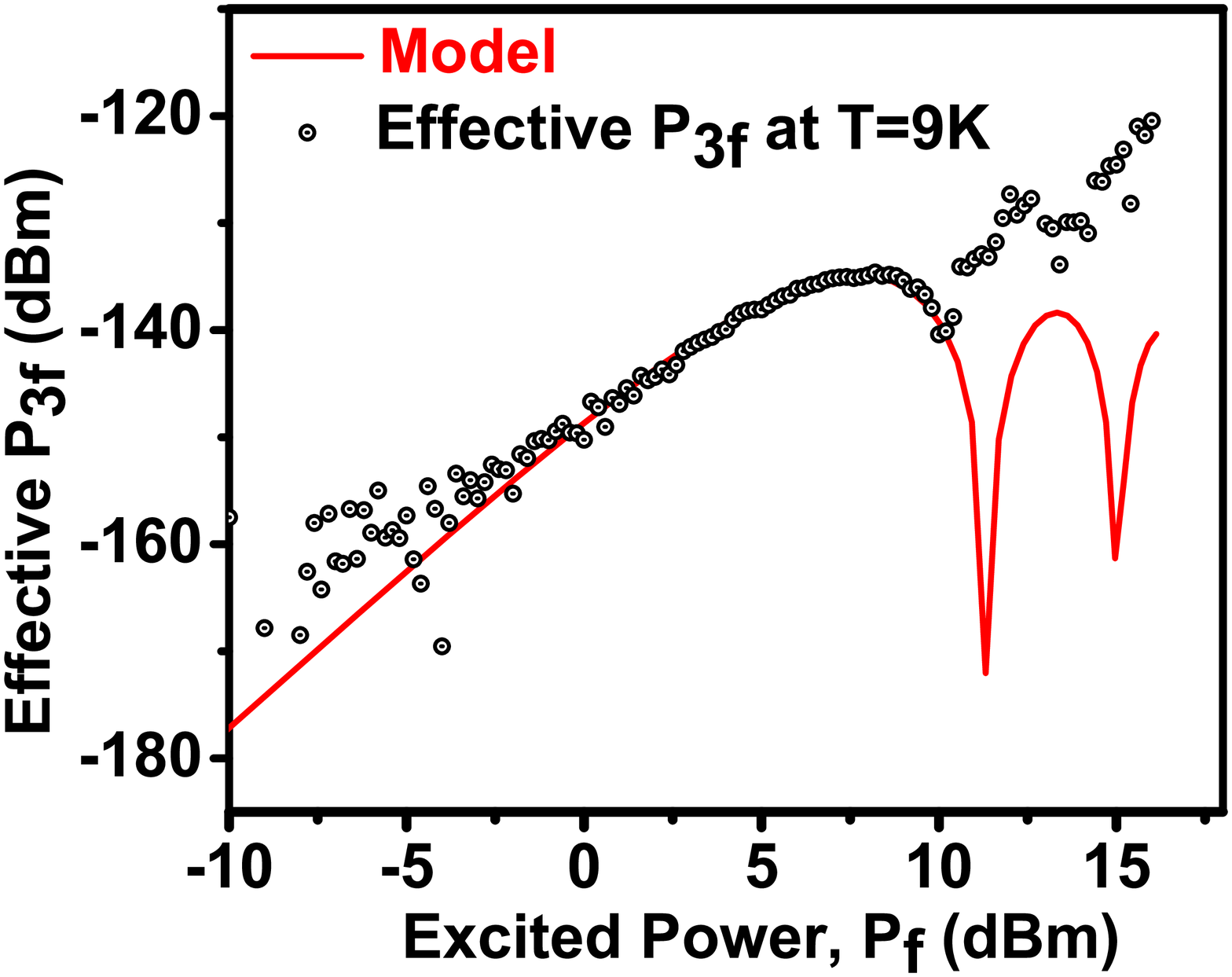}
    \begin{quote}
    \caption{Effective $P_{3f}$ versus $P_f$ for the bulk Nb sample measured by the magnetic write head probe at 9 K under 5.025 GHz excitation as shown by the dot points. The solid line is the theoretical calculation based on the phase slip model of the Josephson junction. The parameters are as follows:  T=9.0 K, Frequency=5.025 GHz, $R_n=90$ $\Omega$, $S=9.11*10^{-14}$ $m^2$, $a=100$ $nm$, N=10, $l_{gap}=100$ $nm$. Only $R_n$ and S were varied to perform the fit.}
    \label{BulkNbP3fPf9K_Fit}
    \end{quote}
    \squeezeup
\end{figure}

The points in Fig. \ref{BulkNbP3fPf9K_Fit} show the effective $P_{3f}$ at the representative temperature of 9 K from the experimental data in Fig. \ref{NbP3fPfSet2}.
The solid line in Fig. \ref{BulkNbP3fPf9K_Fit} indicates the fit to the phase slip model for the temperature of T=9.0 K. The relation between the incident fundamental power, $P_f$, and the actual surface magnetic field experienced on the bulk Nb surface is taken as $P_f  = k {B_{rf}}^2$ with $k=25.6$ $W/T^2$ \cite{Tai2014_APL} directly for fitting the effective $P_{3f}$, where $B_{rf}$ is the amplitude of the surface magnetic field on the bulk Nb superconductor and $k$ is an coupling coefficient between the probe and sample to connect power and magnetic field. The value of $k$ can be judged by the field scale generated by finite element simulator and experimental results on a known conventional superconductor for
calibration \cite{Tai2014_APL}. Again, the fits assume constant values of the probe height $a=100$ $nm$, the number of Josephson junctions $N=10$, and the magnetic gap of the yoke $l_{gap}=100$ nm. Then only the product of $S*B_{rf}$ (which comes from the definition of $\beta\equiv \frac{2 \pi S B_{rf}}{\Phi_0}$) and $R_n$ will affect the calculated $P_{3f}$.

Finally, two parameters, $R_n$ and $S$, are varied to fit the model [Eq. \ref{V3fe}] to the data. Varying $R_n$ can adjust the amplitude of $P_{3f}$ and varying $S$ can laterally shift the theoretical curve to locate the first $P_{3f}\big( P_f \big)$ dip position in the experimental data. From Fig. \ref{BulkNbP3fPf9K_Fit}, the dip position of the model at $P_f=11.3$ dBm and $P_f=15.0$ dBm are matched to dip position of the effective $P_{3f}$ measurement taken at 9 K under 5.025 GHz excitation by taking $S=(320 \: nm)^2$. However, the model shows deep dips but the experimental data does not. It is clear in Fig \ref{NbP3fPfSet2} that the data of $P_{3f}$ at $T <T_c$ intersects with the $P_{3f}^{Probe} (T=10.7 K)$ (background level) at high excitation power, implying deep minima in the harmonic response of the superconductor. The crude background subtraction method from limited data points obscures these minima.  The amplitude of $P_{3f}$ at the first peak is matched to the model by taking $R_n=90 $ $\Omega$. This junction resistance is consistent with Halbritter's estimation for a bulk Nb weak link junction, which is $R_{bl} \approx 10^{-15}$ $\Omega m^2$ \cite{Halbritter1999}\cite{Halbritter1995}. At higher power, the amplitude of the effective $P_{3f}$ does not match the phase slip theoretical curve after the first dip. Several issues may be at play : first, when the applied field is higher than the vortex nucleation field, generation, motion and annihilation of vortices will contribute a significant nonlinear response \cite{Gurevich2008}. In addition, vortices pinned by defects will oscillate with the competition of pinning and image attraction forces, generating harmonic response \cite{Gurevich2008}. This vortex pinning phenomenon becomes significant at strong fields which push the vortex deeper past pinning sites. Another reason is that the probe nonlinearity is over-simplified in the calculation: the phase of the $V_{3f}^{probe}$ is not included when subtracting the measured $P_{3f}$ in the normal state from the measured $P_{3f}$ in the superconducting state.

The loop area $S$ for the fit at T=9 K (Fig. \ref{BulkNbP3fPf9K_Fit}) is $S=9.12*10^{-14}$ $m^2$, equivalent to a circular loop with radius $r=170$ $nm$.  This value of the fit loop radius is on the scale of the magnetic gap ( $\sim$ 100 nm) in the write head probe and implies the resolution of the magnetic write head probe in the near field microwave microscope. This length scale is also consistent with those put forward by Halbritter for the distance between weak links on the surface of air-exposed bulk Nb \cite{Halbritter}\cite{Halbritter1999}\cite{HalbritterSRF}\cite{Halbritter1995}.

\begin{figure}
    \centering
    \includegraphics*[width=3.0in]{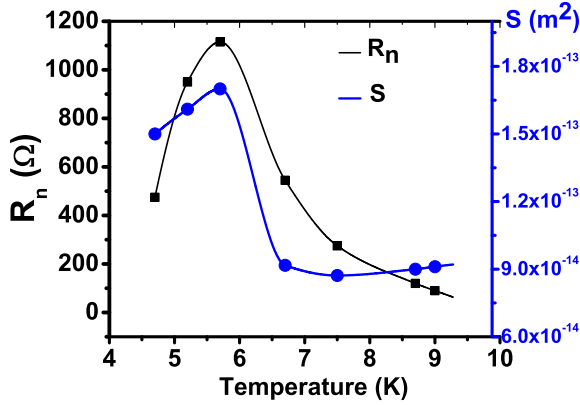}
    \begin{quote}
    \caption{Temperature dependent fitting parameters $R_n$ (normal resistance of the junction) and $S$ (area of the induced screening current loop). Each value is determined by the fitting method of the phase slip model from the $P_{3f}$ versus $P_{f}$ data in Fig. \ref{NbP3fPfSet2}. Note that all data points for both $R_n$ and $S$ are connected by an interpolation spline function.}
    \label{RnS}
    \end{quote}
    \squeezeup
\end{figure}

A summary of the temperature dependent $R_n$ and $S$ from our model fits for the effective $P_{3f}$ at different temperatures is shown in Fig. \ref{RnS}. We see that these two parameters are temperature dependent. (note that the number of junctions in the loop, N, could also vary, but here we choose to keep this value fixed). One reason for temperature dependent $R_n$ and $S$ is because the screening current loop will look for the easiest trajectory with highest critical current of the weak link Josephson junctions to pass through. One can see the $R_n$ gradually goes up with decreasing temperature from 9 K to 5.7 K. This is consistent with the decrease of quasi-particle density upon going to lower temperature. For $S(T)$, a feature of the jump from low-values at T=6.3 K to the larger values at T=5.7 K indicates the loops get bigger at low temperature. The temperature of this jump is coincident with the transition temperature of NbO$_x$ material, or proximity-coupled Nb, in the vicinity of weak links in bulk Nb with reduced transition temperatures of 5.1 K to 6.0 K as identified by Halbritter \cite{HalbritterSRF}. Below 5.7 K, the $R_n(T)$ decreases with decreasing temperature. This decrease also corresponds to a decrease of the area $S$ with decreasing temperature below 5.7 K. Although we don't know how to interpret this decrease of $R_n$ and $S$ at low temperature, several possible issues may be in play. The localized heating of the sample by the probe may be another possible reason for the non-monotonic dependence in Fig. \ref{RnS}. These possibilities may affect the screening current area $S$ at low temperature and result in a corresponding change of $R_n$ at low temperature. These results are consistent with the existence of a dense network of weak links on the surface of air-exposed Nb. It is clear that our microscope is probing this complex network on the scale of the defect size and that the network evolves significantly under RF-field excitation as a function of temperature.

\begin{figure}
    \centering
    \includegraphics*[width=3.0in]{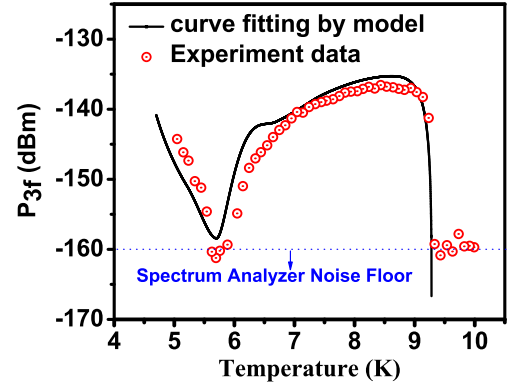}
    \begin{quote}
    \caption{Temperature dependence of the third harmonic response for the
    bulk Nb sample at the excitation power 5.5 dBm and excitation frequency 5.025 GHz measured by the magnetic write head probe (dot points). The noise floor of the spectrum analyzer at this excitation frequency is about -161 dBm. A calculated result (solid curve in black) based on the phase slip model is fit to the data.  The fitting parameter is
    $B_{rf}=11.75$ $mT$, with fixed values for the remaining parameters:
    $l_{gap}$=100 nm, $N=10$, $a=100$ $nm$. }
    \label{BulkNbP3fT_fit}
    \end{quote}
    \squeezeup
\end{figure}

Finally, the experimental temperature dependent $P_{3f}(T)$ with 5.5 dBm excitation in Fig. \ref{BulkNbP3fTset2} is also fitted by the phase slip model. An interpolation function for the temperature dependence $R_n(T)$ and $S(T)$ plotted as the solid line of Fig. \ref{RnS} and used for $P_{3f}(T)$ fitting. The black solid line in Fig. \ref{BulkNbP3fT_fit} is the fit result from the phase slip model with $B_{rf}=11.75$ $mT$. Other parameters ($a=100$ $nm$, N=10, $l_{gap}=100$ $nm$) remain the same as the fitting of $P_{3f}$ on $P_f$. The experimental data points are shown in red circles taken at 5.5 dBm excitation power and 5.025 GHz excitation frequency from Fig. \ref{BulkNbP3fTset2}. Due to the finite number of data points in the interpolation function for the temperature dependence $R_n(T)$ and $S(T)$, the fitting curve shows some discrepancy. However, most of the $P_{3f}(T)$ data points still follow the trend of the phase slip model. The fit of the 0 dBm data in Fig. \ref{BulkNbP3fTset2} is qualitatively correct as well, which makes a compelling case that the nonlinear mechanism comes from the weak link Josephson junction while the excitation power is less than the first deep oscillation in the effective $P_{3f}(P_f)$ curve.

\section{Conclusions}

A clear reproducible nonlinear response signal from the surface of superconducting bulk Nb is obtained by the magnetic write head probe. The success of nonlinear excitation on bulk Nb also implies the magnetic field from the magnetic write head probe is on the order of the thermodynamic critical field of the measured bulk Nb sample. From fits to the data, nonlinearity from the weak-link Josephson effect can explain the measured nonlinear response on the bulk Nb for both $P_{3f}(T)$ and $P_{3f}(P_f)$ measurements. The measured nonlinearity of the bulk Nb samples is perhaps due to the contaminated surface in the fabrication process or subsequent handling in air. The nonlinear near-field magnetic field microwave microscope has great potential to identify the electrodynamic properties of bulk Nb materials and to image the electrodynamic defects on superconducting Nb in the GHz frequency region. Our work in measuring local high frequency microwave properties of Nb superconductors will benefit the SRF community to understand the key issues that lead to degradation in the performance of Nb cavities.

\section{Acknowledgement}
This work is supported by the US Department of Energy $/$ High Energy
Physics through grant $\#$ DE-SC0012036 and CNAM. The work at
MSU was supported by DOE-OHEP, Contract No. DES0004222.


\begin{thebibliography}{99}
%1
\bibitem{ILC}
Abdelhak Djouadi, Joseph Lykken, Klaus M\"{o}nig, Yasuhiro Okada, Mark Oreglia, Satoru Yamashita, et al, ``International Linear Collider Reference Design Report Volume 2: PHYSICS AT THE ILC," arXiv: 0709.1893

%2
\bibitem{H. A. Schwettman}
H. A. Schwettman, J.P. Turneaure and W.M. Fairbank, T.I Smith, M.S.
McAshan, F.B. Wilson, E.E Chamber, ``Low temperature aspects of a
cryogenic accelerator," IEEE Transaction on Nuclear Science
\textbf{14}, (3), 336, PAC (1967).

%3
\bibitem{Padamsee1}
Hasan Padamsee, Jens Knobloch, Tomas Hays ``RF superconductivities on
accelerators (Wiley Series in Beam Physics and Accelerator
Technology), " ISBN: 978-3-527-40842-9, (1998).

%4
\bibitem{Ciovati2010}
G. Ciovati, G. Myneni, F. Stevie, P. Maheshwari, D. Griffis, ``High
field Q slope and the baking effect: Review of recent experimental
results and new data on Nb heat treatments," Phys. Rev. Special
Topics-Accerlerators and Beams \textbf{13}, 022002, (2010).

%5
\bibitem{Sung2011}
Z. H. Sung, A. A. Polyanskii, P. J. Lee, A. Gurevich, D. C. Larbalestier, ``Suppressed Superconductivity on the Surface of Superconducting RF Quality Niobium for Particle Accelerating Cavities," AIP Conf. Proc., \textbf{1352}, 142-150 (2011)

%6
\bibitem{Polyanskii}
A. A. Polyanskii, P. J. Lee, A. Gurevich, Zu-Hawn Sung, D. C. Larbalestier, ``Magneto-Optical Study High-Purity Niobium for Superconducting RF Application," AIP Conf. Proc., \textbf{1352}, 186-202 (2011)

%7
\bibitem{SRF_tutorial2011}
C Antoine, A Aspart, S Regnault, A Chincarini, ``Surface Studies: Method of Analysis and Results," Proceeding of the 10th Workshop on RF Superconductivity, Tsukuba, Japan (2001). $\newline$
CZ Antoine, A Aspart, M Berthelot, Y Gasser, JP Poupeau, F Valin, ``Morphological and Chemical studies of Nb Samples after Various Surface Treatment," Proceeding of the 9th Workshop on RF Superconductivity, 295-303, (1999).

%8
\bibitem{Sung2014}
Z. H Sung, P. J. Lee, D. C. Larbalestier, `` Observation of the Microstructure of Grain Boundary Oxides in Superconducting RF-Quality Niobium With High-Resolution TEM (Transmission Electron Microscope)," IEEE Trans.
Appl. Supercond., \textbf{24}, 68-73 (2014).

%9
\bibitem{HalbritterSRF}
J. Halbritter, ``On Extrinsic - weak link - effects in the surface impedance of cuprate - and classical - superconductors," Fifth Workshop on RF Superconductivity, DESY, Hamburg, Germany (1991).

%10
\bibitem{S. C. Lee1}
S.-C. Lee, S. M. Anlage, ``Study of local nonlinear properties using
a near-field microwave microscope," IEEE Trans. Appl. Supercond.
\textbf{13}, 3594, (2003).

%11
\bibitem{S. C. Lee2}
S.-C. Lee, S. M. Anlage, ``Spatially resolved nonlinearity
measurements of YBa$_{2}$Cu$_{3}$O$_{7}$  bi-crystal grain
boundaries," Appl. Phys. Lett. \textbf{82}, 1893, (2003).

%12
\bibitem{S. C. Lee3}
S.-C. Lee, S.-Y. Lee, S. M. Anlage, ``Microwave nonlinearities of an
isolated long YBa$_{2}$Cu$_{3}$O$_{7-\delta}$  bicrystal grain
boundary," Phys. Rev. B \textbf{72}, 024527, (2005).

%13
\bibitem{Mircea1}
D. I. Mircea and T. W. Clinton, ``Near-field microwave probe for local ferromagnetic
resonance characterization," Appl. Phys. Lett. \textbf{90}, 142504, (2007).

%14
\bibitem{Clinton}
T. W. Clinton, N. Benatmane,J. Hohlfeld, Erol Girt, ``Comparison of a near-field ferromagnetic resonance probe with pumpprobe
characterization of CoCrPt media," J. Appl. Phys. \textbf{103}, (7), 07F546, (2008).

%15
\bibitem{Lamura}
G. Lamura, M. Aurino, A. Andreone, J.-C. Vill\'{e}gier ``First critical field measurements of superconducting films by third harmonic
analysis," J. Appl. Phys, \textbf{106}, 053903, (2009).

%16
\bibitem{Takamasu}
T Takamasu, K Sato, Y Imanaka and K Takehana, ``Fabrication of a micro-coil pulsed magnet system and its
application for solid state physics," Journal of Physics: Conference Series \textbf{51}, 591, (2006).

%17
\bibitem{Bonfin}
K. Mackay, M. Bonfim, D. Givord, and A. Fontaine, ``50 T pulsed magnetic fields in microcoils," J. Appl. Phys. \textbf{87}, 1996, (2000).

%18
\bibitem{Kikuchi}
N. Kikuchi, S. Okamoto, O. Kitakami, ``Generation of nanosecond magnetic pulse field for switching experiments
on a single Co/Pt nanodot," J. Appl. Phys. \textbf{105}, 07D506, (2009).

%19
\bibitem{Woltersdorf}
Georg Woltersdorf, Christian H. Back, ``Microwave assisted switching of single domain $Ni_{80}Fe_{20}$ elements," Phys. Rev. Lett. \textbf{99}, 227207, (2007).

%20
\bibitem{Yoshida}
K. Yoshida., E. Uda, N. Udagawa, and Y. Kanai, ``Investigation on magnetic fields from field-generating layer in MAMR," IEEE Trans. Mag. \textbf{44}, 3408, (2008).

%21
\bibitem{Fang}
D. Fang, H. Kurebayashi, J. Wunderlich, K. V$\dot{y}$orn$\dot{y}$, L. P. Z$\hat{a}$rbo, R. P. Campion, A. Casiraghi,
B. L. Gallagher, T. Jungwirth, A. J. Ferguson1, ``Spin-oebir Driven Ferromagnetic Resonance," Nature Nanotechnology, \textbf{22}, 413, (2011).

%22
\bibitem{Zhu}
X. Zhu, J. G. Zhu, ``Bias-field-free microwave oscillator driven by perpendicularly polarized spin current,"
IEEE Trans. Mag. \textbf{42}, 2670, (2006).

%23
\bibitem{Sim}
C. H. Sim, M. Moneck, T. Liew, J.G. Zhu, ``Frequency-tunable perpendicular spin torque oscillator" J. Appl. Phys. \textbf{111}, 07C914, (2012).

%24
\bibitem{Tai2011}
Tamin Tai, X. X. Xi, C. G. Zhuang, D. I. Mircea, S. M. Anlage,
``Nonlinear near-field microwave microscope for RF defect
localization in superconductors," IEEE Trans. Appl. Supercond.
\textbf{21}, 2615, (2011).

%25
\bibitem{Tai2012}
Tamin Tai, B. G. Ghamsari, T. Tan, C. G. Zhuang, X. X. Xi, Steven M. Anlage, ``MgB$_2$ nonlinear properties investigated under localized high rf magnetic field excitation," Phys. Rev. ST Accel. Beams \textbf{15}, 122002, (2012).

%26
\bibitem{Tai2013}
Tamin Tai, B. G. Ghamsari, Steven M. Anlage, ``Nanoscale electrodynamic response of Nb
superconductors," IEEE Trans. Appl. Supercond. \textbf{23}, 7100104, (2013).

%27
\bibitem{Tai2014_JAP}
Tamin Tai, B. G. Ghamsari, Steven M. Anlage, ``Modeling the nanoscale linear response of superconducting thin films measured by a scanning probe microwave microscope," J. Appl. Phys. \textbf{115}, 203908 (2014).

%28
\bibitem{Tai2014_APL}
Tamin Tai, Behnood G. Ghamsari, Thomas R. Bieler, Teng Tan, X. X. Xi and Steven M. Anlage, ``Near-field microwave magnetic nanoscopy of superconducting radio frequency cavity materials," Appl. Phys. Lett. \textbf{104}, 232603 (2014).

%29
\bibitem{Oates2007}
D. E. Oates, Y. D. Agassi, B. H. Moeckly, ``Intermodulation distortion and nonlinearity
in $MgB_2$: experiment and theory," IEEE Trans. Appl. Supercond. \textbf{17}, (2), 2871-2874, (2007).

%30
\bibitem{Booth1999}
J. C. Booth, J. A. Beall, D. A. Rudman, L. R. Vale, R. H. Ono, ``Geometry dependence of nonlinear effects in high temperature superconducting transmission lines at microwave frequencies," J. Appl. Phys. \textbf{86}, 1020, (1999).

%31
\bibitem{Bieler}
Di Kang, Derek C. Baars, Thomas R. Bieler, and Chris C. Compton, ``Characterization of large grain Nb ingot microstructure using EBSP mapping and Laue camera methods," AIP Conf. Proc. \textbf{1352}, 90-99, (2011).

%32
\bibitem{Tai_thesis}
Tamin Tai, ``Measuring electromagnetic properties of superconductors in high and localized RF magnetic field," Ph.D. dissertation, University of Maryland-College Park, (2013), see http://hdl.handle.net/1903/14668.

%33
\bibitem{Carson Jeffries}
Carson Jeffries, Q. Harry Lam, Youngtae Kim, L. C. Bourne, and A. Zettl, ``Symmetry breaking and nonlinear electrodynamics in the ceramic superconductor $YBa_2Cu_3O_7$,"  Phys. Rev. B. \textbf{37}, 9840, (1988).

%34
\bibitem{Xia and Stround}
Ting-kang Xia, D. Stroud, ``Nonlinear electrodynamics and nonresonsnt microwave absoption in ceramic superconductors," Phys. Rev. B. \textbf{39}, (7), 4792, (1989).

%35
\bibitem{Halbritter}
J. Halbitter, ``On the oxidation and on the superconductivity of niobium," Appl. Phys. A-Mater. \textbf{43}, (1), 1-28 (1987).

%36
\bibitem{Ciovati2006}
G. Ciovati and J. Halbritter, ``Analysis of the medium field Q-slope in superconducting cavities made of bulk niobium," Physica C: Superconductivity \textbf{441}, (1-2), 57-61 (2006).

%37
\bibitem{Gurevich2008}
A. Gurevich, G. Ciovati, ``Dynamics of vortex penetration, jumpwise
instabilities, and nonlinear surface resistance of type-II
superconductors in strong rf fields," Phys. Rev. B \textbf{77},
104501, (2008).

%43
\bibitem{Tinkham_Book}
Michael Tinkham, ``Introduction to superconductivity," 2nd Edition, ISBN-10: 0486435032 | ISBN-13:978-0-486-43503-9, (2004).

%38
\bibitem{Peeters2004}
M. V. Milo$\check{s}$evi$\acute{c}$, F. M. Peeters, ``Vortex pinning in a superconducting film due to in-plane magnetized ferromagnets of different shapes: The London approximation," Phys. Rev. B. \textbf{69}, 104522, (2004).

%39
\bibitem{Peeters2011}
A. V. Kapra, V. R. Misko, D. Y. Vodolazov and F. M. Peeters, ``The guidance of vortex-antivortex pairs by in-plane magnetic dipoles in a superconducting finite-size film," Supercond. Sci. Technol. \textbf{24}, 024014, (2011).

%40
\bibitem{Bael}
M. J. Van Bael, J. Bekaert, K. Temst, L. Van Look, M. V. Moshchalkov, Y. Bruynseraede, G. D. Howells, A. N. Grigorenko, S. J. Bending ``Local observation of field polarity dependent flux pinning by magnetic dipoles," Phys. Rev. Lett. \textbf{86}, (1), 155-158, (2001)

%41
\bibitem{Peeters}
M. V. Milo$\check{s}$evi$\acute{c}$, S. V. Yampolskii, and F. M.
Peeters, ``Magnetic pinning of vortices in a superconducting film:
The antivortex magnetic dipole interaction energy in the London
approximation," Phys. Rev. B \textbf{66}, 174519, (2002).

%42
\bibitem{Ambegaokar and Baratoff}
V. Ambegaokar and Baratoff, ``Tunneling between superconductors," Phys. Rev. Lett. \textbf{10}, 486, (1963)

%44
\bibitem{Waldram}
J. R. Waldram, ``Superconductivity of metals and cuprates," Chapter \textbf{10}. (The non-local form of BCS theory)  ISBN-10: 0852743378 | ISBN-13: 978-0852743379, (1996).

%45
\bibitem{Halbritter1999}
J. Halbritter, "Materials science and surface impedance Z (T,f,H) of Nb and YBCO and their quantitative modelling by the leakage current of weak links," Supercond. Sci. Tech. \textbf{12}, (11), 883-886, (1999).

%46
\bibitem{Klaas B. Klaassen}
Klaas B. Klaassen, Jack C.L. van Peppen, ``Nanosecond and sub-nanosecond writing experiments," IEEE Trans. Magn. \textbf{35}, No.2, 625, (1999).

%47
\bibitem{ReadRiteCorp}
F. H. Liu, S. Shi, J. Wang, Y. Chen, K.
Stoev, L. Leal, R. Saha, H.C. Tong, S. Dey, M. Nojaba, ``Magnetic
recording at a data rate of one gigabit per second," IEEE
Trans. Appl. \textbf{37}, (2), 613-618, (2001).

%48
\bibitem{Halbritter1995}
J. Halbritter, ``Granular superconductors and their intrinsic and extrinsic surface impedance," Journal of Superconductivity \textbf{8}, (6), 691-703, (1995).
\end{thebibliography}
\end{document}